\documentclass[10pt]{article}
\usepackage[utf8]{inputenc}
\usepackage[english]{babel}
\usepackage[round]{natbib}
\usepackage{subcaption}
\usepackage{graphicx}
\usepackage{amsmath}
\usepackage{amssymb}
\usepackage[unicode]{hyperref}
\usepackage[tableposition=top]{caption}
\usepackage{array}
\usepackage{multirow}
\usepackage{tabularx}
\usepackage{textcomp}
\usepackage{gensymb}
\usepackage{graphics}
\usepackage{color}
\usepackage{cleveref}
\usepackage[a4paper, total={6.5in, 8.5in}]{geometry}
\numberwithin{figure}{section}
\numberwithin{table}{section}
\usepackage{titlesec}
\titleformat{\subsection}{\normalfont\itshape}{\thesubsection}{1em}{}
\usepackage{chngcntr}
\counterwithout{figure}{section}
\counterwithout{table}{section}

\title{Tokamak disruption prediction using different machine learning techniques}
\author{J. Croonen\footnote{Correspondence email: joost.croonen@kuleuven.be}, J. Amaya and G. Lapenta \\ \small{Centre for mathematical Plasma Astrophysics, KU Leuven, Leuven, 3000, Belgium}}
\date{May 11, 2020}

\begin{document}

\maketitle
%TODO: github
%TODO: refs
%TODO: structure
\begin{abstract}
    Disruption prediction and mitigation is of key importance in the development of sustainable tokamak reactors. Machine learning has become a key tool in this endeavour. In this paper multiple machine learning models will be tested and compared. A particular focus has been placed on their portability. This describes how easily the models can be used with data from new devices. The methods used in this paper are support vector machine, 2-tiered support vector machine, random forest, gradient boosted trees and long-short term memory. The results show that the support vector machine performance is marginally better among the standard models, while the gradient boosted trees performed the worst. The portable variant of each model had lower performance. Random forest obtained the highest portable performance. Results also suggest that disruptions can be detected as early as 600ms before the event. An analysis of the computational cost showed all models run in less than 1ms, allowing sufficient time for disruption mitigation.
\end{abstract}

\section{Introduction}
Disruptions are highly energetic events in tokamak fusion devices, which result in loss of plasma confinement and severe thermal and mechanical stress on device components \citep{Boozer, Cardella}. As machines are scaled to reactor sizes, the damage such events can cause can be catastrophic. Disruptions therefore pose a serious threat and avoiding them is a high priority goal in the further development of the technology \citep{ITER_DMW}. In the last two decades machine learning has become a promising tool in attaining this goal. It can be used to predict the disruptions in advance such that their effects can be mitigated \citep{Baylor, Jachmich}. These methods don't rely on a first principle understanding of the plasma and disruption physics. Instead they use the vast amount of experimental data available to train models to detect patterns associated with disruptions. Such methods have been successfully applied in several devices around the world like JET and ASDEX-Upgrade \citep{Moreno, Cannas}. In this paper, based on \citet{my_thesis}, a comparative study is performed between different machine learning methods: support vector machines, 2-tiered support vector machines, random forest, gradient boosted trees and long-short term memory. The models were not designed to compete against existing predictors, but were designed to allow comparisons among them. This gives an insight into the benefits and drawbacks of each individual method. Particular attention has been put on the portability of the different methods: how easy it is to transfer a model, trained on one tokamak, to a new device. Such models would be particularly useful for the next generation of tokamaks, like ITER, since there is no experimental data available to train predictor models for such devices. 

The following topics will be discussed in this paper: Section \ref{sec:data} discusses the data and normalisation techniques that were used for the training of the models. Section \ref{sec:methods} describes the machine learning methods used in this work. Section \ref{sec:results} discusses the performance metrics and results of the ML models. Section \ref{sec:conclusion} summarises the most important results of this paper. The code used to train the models and analyse performance is available at \url{https://github.com/JoostCroonen/ML\_Tokamak\_Disruption\_Prediction}.

\section{Data}\label{sec:data}
The different Machine Learning (ML) algorithms were trained using data from the Joint European Torus (JET), governed by the Culham Centre for Fusion Energy (CCFE). Disruption events in this dataset are recorded in the ITPA (international tokamak physics activity) disruption database (IDDB)\citep{iddb}. IDDB contains shot numbers and times of disruption for 1170 disruptive events in JET. The disruption range between shots 32157 and 79831. They do not contain any artificially induced disruptions. The dataset is split into a training set (60\%), used to train the models, a test set (20\%) to determine the performance and a validation set (20\%) which is used to determine the hyperparameters (see section \ref{sec:methods}).

\begin{table}
    \centering
    \setlength\tabcolsep{25pt}
    \renewcommand{\arraystretch}{1.2}
    \caption{List of the input features used to train the different predictor models. Selection based on \citet{ratta_2012} and \citet{Rea}.}
    \label{tab:feat}
    \begin{tabular}{ll}
    \hline
        name            & description \\
        \hline
        $I_{pla}$       & plasma current \\
        $MLA $          & mode lock amplitude \\
        $l$             & plasma internal inductance \\
        $W_{dia}$       & diamagnetic energy \\
        $\dot{W_{dia}}$ & time derivative of the diamagnetic energy \\
        $n_e$           & electron density \\
        $P_{out}$       & radiated output power \\
        $P_{in}$        & input power: sum of ICRH and NBI power \\
        $q_{95}$        & edge safety factor \\
        $B_{\phi}$      & toroidal magnetic field strength \\
    \hline
    \end{tabular}
\end{table}

\begin{figure}
    \centering
    \includegraphics[scale=0.5]{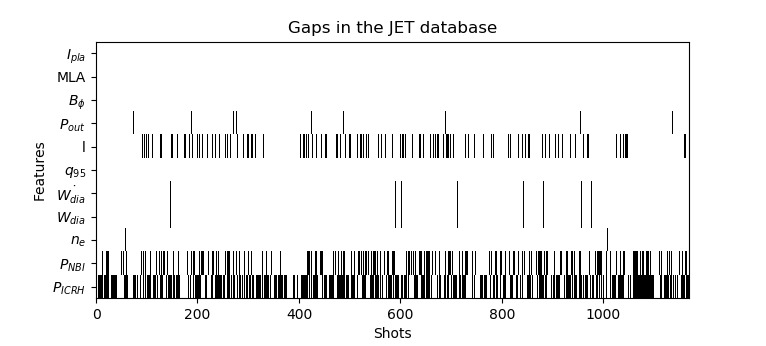}
    \caption{Visualisation of missing data from the JET database for the relevant shots. Black lines represent a gap for the corresponding shot and input feature.}
    \label{fig:gaps}
\end{figure}

The input signals for the ML models, also called features, are listed in table \ref{tab:feat}. These features were chosen based on work by \citet{ratta_2012} and \citet{Rea}. They were normalised around 0 with standard deviation of 1. The sampling cadence of all the signals were locked to 30ms, the cadence of the majority of the signals. 
Some signals are unavailable for specific shots. These gaps in the database are visualised in figure \ref{fig:gaps}, where the black lines correspond to missing data. The incomplete dataset makes it more difficult for ML models to learn from the available data. In particular the neutral beam injector (NBI) power and ion-cyclotron radio heating (ICRH) power are frequently missing. Due to the relatively small dataset, it was decided best to keep all shots, even if data was missing from them. The missing data was set to zero. 

The samples consisted of several 600ms sequences (equivalent to 20 datapoints) for each shot. Disruptive (positive) samples are taken from the range of 30-630ms before disruption. The final 30ms is the minimum time required for disruption mitigation systems to be activated in time. 
Non disruptive samples (negative) are taken from 1000ms-1600ms and 4000ms-4600ms before disruption. 
Here the assumption is made that at sufficiently long timescales before a disruption the disruptive nature of a shot should not yet be discernible. The exact time before disruption where this becomes true is difficult to estimate. The choice here is based on \citet{Aledda}, were this was set to 1s before disruption.

\section{Methods}\label{sec:methods}
ML is a useful tool for the prediction of disruption. It does not require a thorough understanding of the mechanisms behind these events. ML makes use of the large amount of experimental data that is available from experimental reactors, to detect patterns associated with disruptions. This information is used to classify new samples as either disruptive or non-disruptive. 

ML algorithms non-linearly process the inputs using a large number of parameters which determine the output. To train the models, the error between the calculated output and the desired output is calculated using a cost function. This cost function is subsequently minimised using gradient descent. This means the parameters are adjusted according to the gradient of the cost function with respect to these parameters. 60\% of the dataset was used for training. 

Each of the algorithms used in this work has a set of hyperparameters determining their learning behaviour. These have been optimised for the validation set (20\% of dataset) through a grid search method. The performance values are measured using the test set, consisting of the final 20\% of the dataset.  
The validation set is distinct from the test set to avoid tailor-fitting the hyperparameters to the test set, which would unrealistically inflate performance results. 

All the models were trained twice: once as a standard predictor and once as a portable predictor. The latter means the model can be used across different tokamaks. This allows a ML models trained on an older device for which there is a lot data available, to be used on a new tokamak for which there is no data available. To design the portable predictor models the input signals were normalised to be machine independent. This was done as shown in table \ref{tab:port_feat}. These portable features were chosen based on \citet{tang} and \citet{Rea}. 

The code for training and analysing the different models is available at \url{https://github.com/JoostCroonen/ML\_Tokamak\_Disruption\_Prediction}.

\begin{table}
    \centering
    \setlength\tabcolsep{25pt}
    \renewcommand{\arraystretch}{1.2}
    \caption{List of machine independent features that are used to train the portable predictors and how they are derived from the original set of features in table \ref{tab:feat}}
    \label{tab:port_feat}
    \begin{tabular}{lll}
    \hline
        name            & description                   & formula\\
        \hline
        $I_{N}$         & normalized plasma current     & $I_{pla}/B_\phi$ \\
        $f_{MLA} $      & mode lock amplitude fraction  & $MLA/(aB_\phi)$\\
        $l$             & plasma internal inductance    & $l$\\
        $f_{gw}$        & Greenwald density fraction    & $n_e/(I_{pla}/\pi a^2)$\\
        $f_{P}$         & radiated power fraction       & $P_{out}/(P_{in} - \dot{W_{dia}})$\\
        $q_{95}$        & edge safety factor            & $q_{95}$\\
    \hline
    \end{tabular}
\end{table}

\subsection{Support vector machines}
The first algorithm is the support vector machine (SVM). It is designed to find the hyperplane that best separates two distinct classes in feature space. In this problem the classes are disruptive and non disruptive and feature space has as dimensions the different input signals of the SVM. It decides on the best hyperplane by maximising the margin around it. 
When a dataset cannot be linearly separated, a transformation function can be used to map the data into a higher dimensional space where the datapoints are linearly separable, which is known as the kernel trick. A detailed description and mathematical derivation of this model can be found in \citet[Chapter 12]{ML}.

\begin{figure}
  \centering
  \includegraphics[scale=0.4]{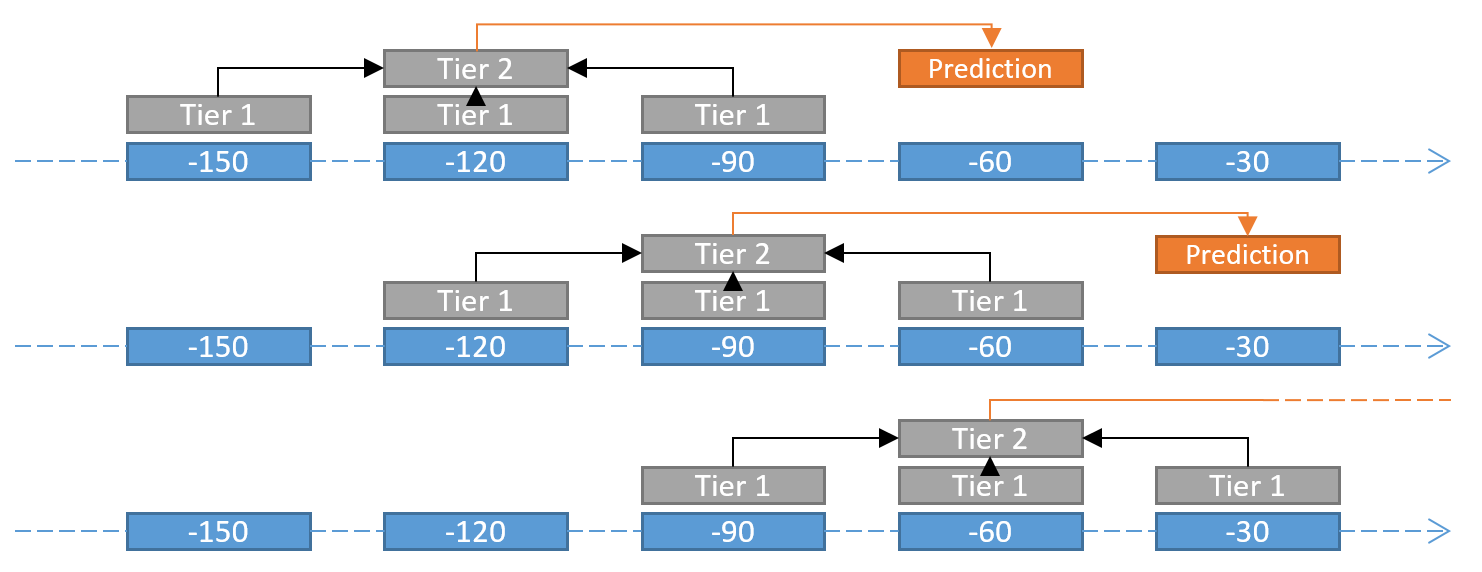}% Images in 100% size
  \caption{Schematic of the sliding window scheme used by the 2-tiered SVM. (Source: \citet{my_thesis})}
\label{fig:2t_slide}
\end{figure}

The simplicity of the model and low computational cost makes it an excellent tool for quick iteration and testing. It has also been used in previous work like \citet{Moreno}. 
It was implemented using the python package Scikit-learn \citep{scikit-learn}. It uses the radial basis function (RBF) as kernel with regularisation parameter $C = 0.33$ and kernel size $\gamma = 0.1$. 
Based on \citet{Moreno} a 2-tiered SVM classifier was also designed. It works based on a sliding window scheme where the first tier SVMs will observer 3 subsequent timesteps. The results of the 3 first tiers is then used by a second tier SVM which outputs the final classification. A schematic of this is shown in figure \ref{fig:2t_slide}. The first tier is identical to the standard SVM except that it outputs probabilities rather than discrete classification while the second tier uses a linear kernel and $C = 0.01$. Due to the implementation in Scikit-learn the two tiers were trained independently. This means errors could not propagate from the second into the first tier SVMs, which could theoretically improve performance. 

\subsection{Random forest}
Random forest (RF) solves classification problems using many different decision trees (DT). DTs are tree like data structures where each branch point represents a decision point. There, one or more variables can be evaluated and based on the outcome, the process will continue along the corresponding branch to the next branch point. This continues until it reaches one of the end points of the tree, known as the leaves. However, a single DT cannot represent the complex behaviour of disruptions.
Combining the outputs of many DTs through a majority voting system, makes for a much more robust classification model. 
The influence of each DT towards the final decision is adjusted during the training process based on their individual performance. A detailed discussion about RF and their training can be found in \citet[Chapter 15]{ML}. RF was also implemented using Scikit-learn. It used $1000$ initial DTs. The RF method allows for analysis of the relative importance of the different inputs. A discussion on how this works can be found in \citet[pp. 367-369, 593]{ML}.

\subsection{Gradient boosted trees}
Boosting algorithms are a group of iterative approximation techniques. On each iteration a new function is introduced whose parameters are adjusted to approximate the residual error from the previous step. In the case of gradient boosted trees (GBT) the estimator functions in each step are DTs. On the first step the residual error is equal to the value of the expected outputs. As the algorithm progresses it will approximate the desired output because each DT corrects for the error from the previous step. 
More details on the working and training of this method can be found in \citet[pp. 342-343]{ML}. 
This method was also implemented with Scikit-learn and used $1000$ iterations.

\subsection{Long short-term memory}
Long short-term memory (LSTM) is a variation of a recurrent neural network (RNN). A RNN is a neural network (NN) taking as an additional input the output signals of the NN at the previous timestep. This ensures there is a link to past events uncovering time dependent behaviour. LSTM improves on this algorithm by introducing a 'memory' called the cell-state. Only read and write operations are applied to this cell-state during training, allowing the NN to maintain information over a long time. 

\begin{figure}
  \centering
  \includegraphics[scale=0.5]{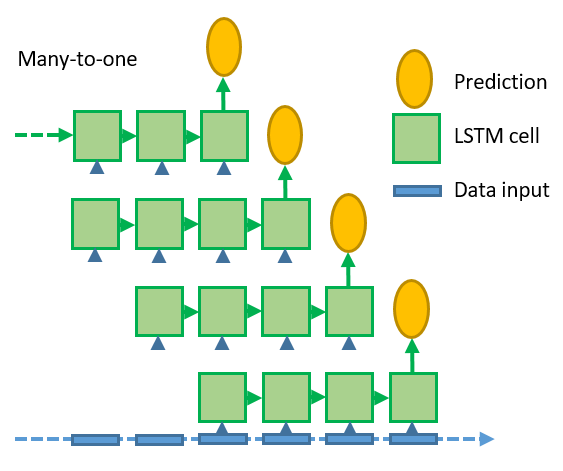}% Images in 100% size
  \caption{Schematic of the sliding many-to-one implementation of LSTM. (Source: \citet{my_thesis})}
\label{fig:m2o}
\end{figure}

A more detailed discussion on the working and training of LSTM can be found in \citet[pp. 410-411]{DL}. The LSTM implementation used in this work is a many-to-one network. It works like a sliding window scheme. On each timestep, the algorithm looks at the last 20 timesteps, on which it performs the LSTM algorithm to make a prediction. This is schematically shown in figure \ref{fig:m2o}. This method ensures good generalisability because it is agnostic of the total sequence length, which can vary significantly between shots. However, it is limited to information of the last 20 timesteps, corresponding to a 600ms time period. Implementation of this method was done using pytorch \citep{torch}. The standard model used a two layer LSTM with 22 nodes while the portable model used a one layer LSTM with 18 nodes. 

Due to the random initialisation of the LSTM, different runs can give different results. Therefore it is not guaranteed that these are the optimal parameters, especially since the difference in performance during hyperparameter fitting was very small. 
A single fully connected layer was used to reduce the final LSTM output to a single value. The model was trained using the ADAM optimiser \citep{Kingma} and using a binary cross-entropy error function. It was trained for 1000 cycles on a single GPU using stochastic mini-batching to reduce the memory footprint.

\section{Results}\label{sec:results}
Commonly used metrics in the field of ML are used to discuss the performance of the models in this work. In a predictive model with binary selection the recall ($R$) is the ratio of the true positives (tp, i.e. the samples that were correctly identified as positive) over the actual positives (ap, i.e. the samples that are actual disruptions). The precision ($P$) is the ratio of the true positives over the predicted positives (pp, i.e. the samples which are predicted by the model, correctly or incorrectly, to be positive). 
\[R = \frac{tp}{ap}\]
\[P = \frac{tp}{pp}\]
The recall is therefore an estimate of the fraction of disruptions that have been correctly predicted by the model. The precision estimates the chance that a prediction is correct when it predicts a disruption. Ideally both should be close to 1. A combined metric, called the F1-score, mixes these two metrics:
\[F1 = \frac{2RP}{R+P}\]
Since the severe consequences of disruptive events it could be advantageous to prioritise the optimisation of the recall. However, a low precision would cause disruption mitigation techniques to be activated unnecessarily decreasing the performance of the fusion device. In this work the two metrics were chosen to be equally important. A sequence of 600ms is labelled positive if at least one timestep is classified as disruptive and negative otherwise. A predicted positive is classified as a true positive if a disruption is imminent within 600ms i.e. if it is in the sequence before a disruption. 

\begin{figure}
  \centering
  \includegraphics[scale=0.5]{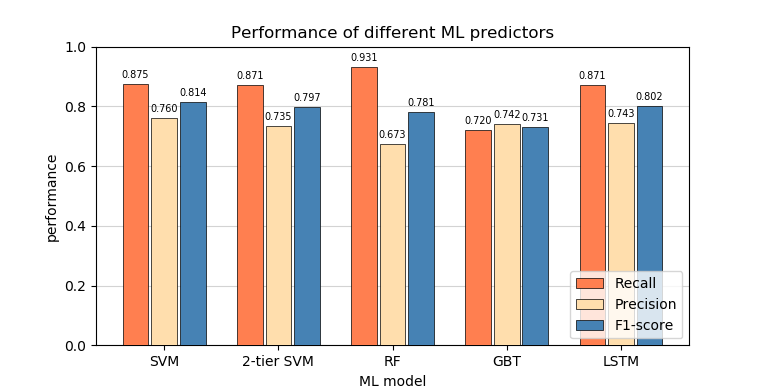}% Images in 100% size
  \caption{Recall, precision and F1-score of the different standard prediction models.}
\label{fig:perf}
\end{figure}

\subsection{Predictor performance}
The performance metrics of the different techniques are shown in figure \ref{fig:perf}. It is apparent that most algorithms prefer optimising the recall. This suggests that for this dataset it might be easier to predict false positives. RF has the strongest imbalance towards recall. GBTs are the exception, with a very balanced performance. The F1 scores of the different models are very similar. SVM has the best overall performance, while GBT has the lowest. Interestingly the 2-tiered SVM does not outperform the standard SVM. Considering the first tier in this model is nearly identical to the standard SVM, it is remarkable that the model shows worse performance. The LSTM also under performed compared to expectations. It was expected that due to its ability to take into account multiple timesteps it would have an advantage over the other models. However, it still only performs similarly. 

\begin{figure}
  \centering
  \includegraphics[scale=0.5]{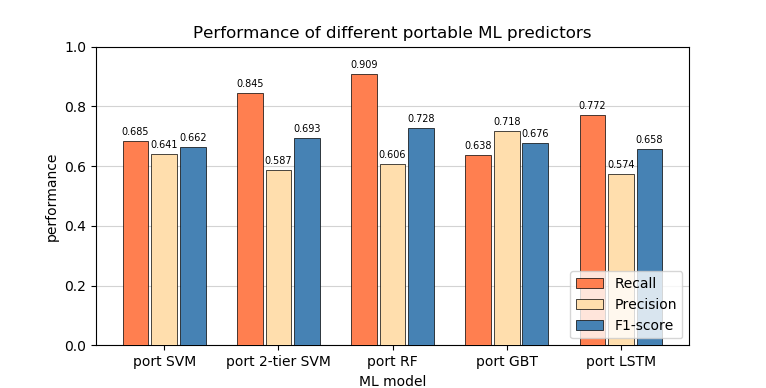}% Images in 100% size
  \caption{Recall, precision and F1-score of the different portable prediction models.}
\label{fig:port_perf}
\end{figure}

Figure \ref{fig:port_perf} shows portable predictor performance metrics. There is an overall loss in performance across the board. This can most likely be attributed to the loss in information in the transition from standard to portable inputs. In the latter the 10 original inputs are combined into only 6 features. This implies is a loss of information, resulting in decreased performance. However it can be seen that not all ML methods are equally sensitive to this transition. SVM and LSTM, which had the best performance in the standard models, had a drop of around $0.15$ in F1-score. The 2-tiered model only dropped about $0.10$. GBT and RF experienced the lowest impact with a drop of only $0.05$. This makes RF the best performing among the portable models. 
These values are only an indication of the upper limit of the portable performance since they were both tested and trained on JET data.

\begin{figure}
\centering
\includegraphics[scale=0.5]{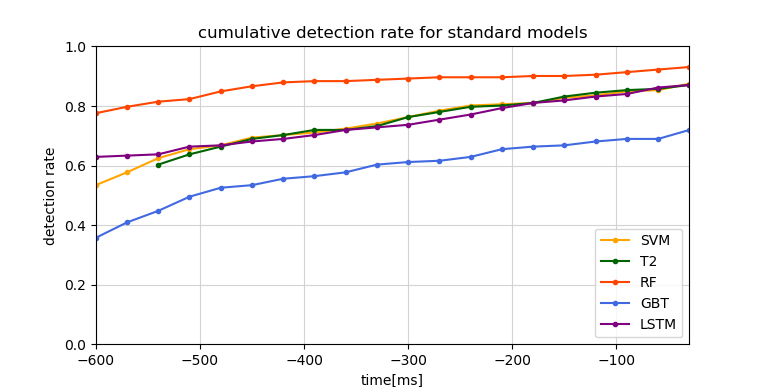}% Images in 100% size
\caption{Cumulative detection rate for the different standard predictor models.}
\label{fig:ca_std}
\end{figure}

\begin{figure}
  \centering
  \includegraphics[scale=0.5]{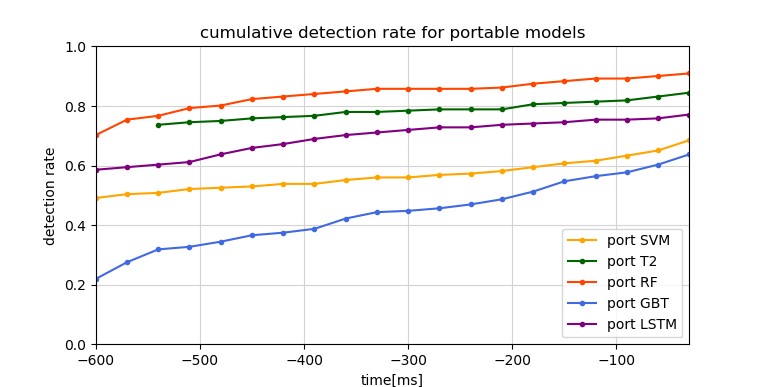}% Images in 100% size
  \caption{Cumulative detection rate for the different portable predictor models.}
\label{fig:ca_port}
\end{figure}

\subsection{Time evolution of the recall}
Figure \ref{fig:ca_std} and \ref{fig:ca_port} show the evolution of the recall in time before disruption for the standard and portable models. It shows when disruptions are detected. The right-most value in these graphs are the final reported recall values. It shows at what times disruptions are being detected. For all the models the recall starts at a high value. This indicates a significant fraction of the disruptions are already detectable at -600ms, suggesting that a longer time window before the disruption should be considered. 

\subsection{Relative feature importance}
\begin{figure}
  \centering
  \includegraphics[scale=0.5]{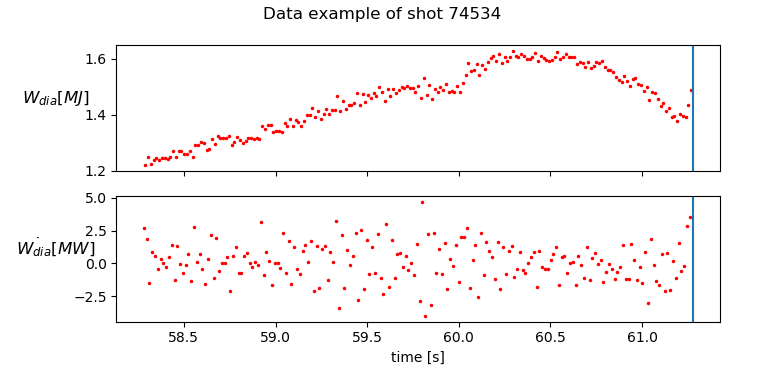}% Images in 100% size
  \caption{Example of diamagnetic energy and its time derivative, showing the noise on the signals. The vertical line is the moment of disruption.}
\label{fig:noise}
\end{figure}

\begin{figure}
    \centering
    \includegraphics[scale=0.5]{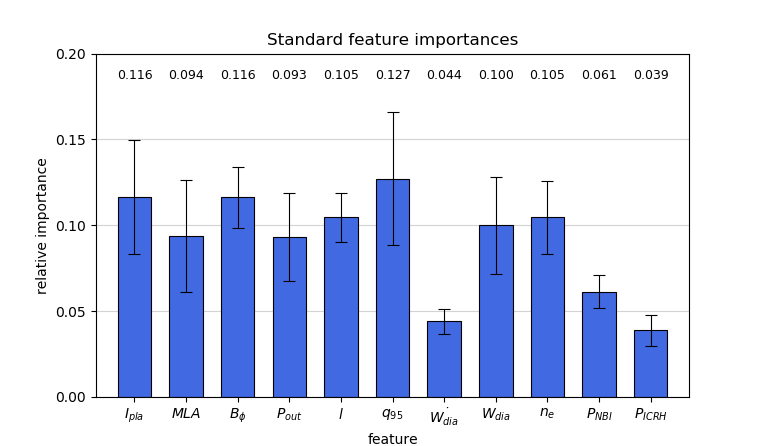}
    \caption{Relative importance of the different standard input features.}
    \label{fig:feat_imp}
\end{figure}

\begin{figure}
    \centering
    \includegraphics[scale=0.5]{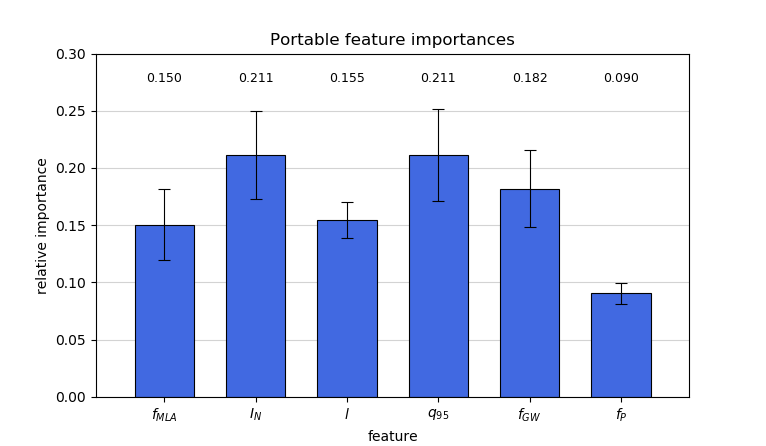}
    \caption{Relative importance of the different portable input features.}
    \label{fig:feat_imp_port}
\end{figure}

The RF method has been used to determine relative importance of the different input features. Results are shown in figure \ref{fig:feat_imp} for the standard models. The time derivative of the diamagnetic energy has the lowest importance of all inputs. This could be due to the noise on the diamagnetic energy signal which can be seen in figure \ref{fig:noise}. This makes for a very erratic time derivative. The ICRH and NBI input power also have a low importance. This is likely caused by the frequent absence of this data in the database. Figure \ref{fig:feat_imp_port} contains the feature importance's for the portable models. Here the radiated power fraction $f_P$ has the lowest importance. $f_P$ depends on $\dot{W_{dia}}$ and $P_{in} = P_{NBI} + P_{ICRH}$, which have the lowest importance in the standard model as well. 

\subsection{Computational cost}
Table \ref{tab:time} shows both train and inference time for the different models. These values are hardware dependent. In this project an intel i7-7700HQ at 2.8GHz was used. The LSTM model also used an Nvidia GTX 1050ti 4GB GPU as accelerator during training, significantly speeding up training time to values similar to the other models. GBT and SVM are the fastest in terms of training time, followed by RF, LSTM and lastly the 2 tiered SVM. The first tier of this model needs to output probabilities of disruption rather than discrete classifications, which significantly increases the training time. 
The inference time is the average time required to make a new prediction, based on the test data. GBT is the fastest, followed by LSTM and RF. Finally there is SVM and the 2 tiered SVM. The tiered model is the slowest since it requires two SVMs to be evaluated. However, the second tier SVM is much simpler and therefore only slightly increases compute time over the standard SVM. Even the slowest inference times are still within 1ms. Slower inference times mean there could be insufficient time for disruption mitigation systems to be activated. 

\begin{table}
    \centering
    \setlength\tabcolsep{15pt}
    \renewcommand{\arraystretch}{1.2}
    \caption{Train and inference time information for the different models. (* LSTM was trained on a GPU to speed up computation)}
    \label{tab:time}
    \begin{tabular}{llllll}
    \hline
            & SVM & T2 & RF & GBT & LSTM  \\
        train time [s] & $28.21$ & $184.89$ & $138.52$ & $24.88$ & $79.18$* \\
        inference time [ms] & $0.43$ & $0.52$ & $0.11$ & $0.007$ & $0.05$ \\
    \hline
    \end{tabular}
\end{table}

\section{Conclusion}\label{sec:conclusion}
Five disruption predictors have been designed based on different machine learning techniques. The performance of the different models, are markedly similar. Nonetheless SVM produces the best results. The 2-tiered variant of the SVM performed worse than the original. The LSTM model was expected to perform better than its competitors, but results in this work don't show an advantage over SVMs. GBT had the lowest performance. The portable models, using machine independent inputs, had overall lower performance. This was expected due to the loss of information in the conversion to portable input features. However, some models were less susceptible to this degradation of performance. In particular RF and GBT maintained there performance much better. Overall this made RF the best portable predictor. The cumulative detection rate of the different models suggest time sequences longer than 600ms should have been considered, since the detection rate at 600ms before disruption was already very high. The train time of the different models showed SVM and GBT to be the fastest to train. LSTM had the highest computational cost, but this was mitigated by training on a GPU, leaving the 2-tiered SVM as the slowest to train. At runtime LSTM and GBT were the fastest, but even the slowest models were adequately fast to work in real time. 

\section*{Acknowledgements}
The authors would like to thank EUROfusion and CCFE for access to the JET database. Special thanks to dr. Dirk Van Eester from LPP-ERM/KMS for his support and invaluable help in accessing and using this database. 
The work reported here received support from the Onderzoekfonds KU Leuven (project C14/19/089). This research used resources of the National Energy Research Scientific Computing Center, which is supported by the Office of Science of the US Department of Energy under Contract No. DE-AC02-05CH11231.

\bibliographystyle{plainnat}
\bibliography{ref}

\end{document}